\documentclass[twocolumn,showpacs,preprintnumbers,amsmath,amssymb]{revtex4}
\usepackage{graphicx}
\usepackage{dcolumn}
\usepackage{bm}
\usepackage[dvips]{color}

\begin{document}


\title{Engineering Entanglement: The Fast-Approach Phase Gate}
\author{Dan Vager, Bilha Segev\footnote{Professor Bilha Segev died on
March 17 of complications arising in her battle with cancer.  We
lost a very dear colleague; a colleague who shared her enthusiasm
for science and for life with us.  She will be sorely missed.},
and Y. B. Band} \affiliation {Ben-Gurion University of the Negev,
Beer-Sheva 84105, Israel}

\date{\today}

\begin{abstract}
Optimal-control techniques and a fast-approach scheme are used to
implement a collisional control phase gate in a model of cold
atoms in an optical lattice, significantly reducing the gate time
as compared to adiabatic evolution while maintaining high
fidelity.  New objective functionals are given for which optimal
paths are obtained for evolution that yields a control-phase gate
up to single-atom Rabi shifts.  Furthermore, the fast-approach
procedure is used to design a path to significantly increase the
fidelity of non-adiabatic transport in a recent experiment.  Also,
the entanglement power of phase gates is quantified.
\end{abstract}

\pacs{03.67.-a, 03.67.Mn, 03.65.-w,  03.65.Ca} \maketitle

\section{Introduction}
Quantum information processing with cold atoms in optical lattices
\cite{Jaksch_98, deutsch98, Brennen_99, Jaksch_99, Calarco,
Tiesinga, 1dim} and in micro-traps \cite{Jaksch_99,
Calarco,Eckert_02, Dumke_02} rely on the ability to entangle
nearest neighbor atoms in an efficient controlled way.  The
effective trapping potential in optical lattices can have the form
of a double well where the distance between minima, the well
separation height, the frequency of each well, etc., can each be
manipulated by controlling the lasers, using different
combinations of laser frequencies and bias fields
\cite{Guidoni_99}.  Ideally each atom is initially in one of the
optical wells, and single qubits are registered into each atom's
state.  The atom's internal states (e.g., hyperfine levels)
\cite{Brennen_99, Jaksch_99} or motional states in the trap
\cite{Eckert_02, Tiesinga} can be used as a computational basis,
$|n\rangle$, where $n=0,1$.  A two-qubit gate is implemented by
bringing the atoms together and letting their wave functions
overlap \cite{Brennen_99, Jaksch_99, Calarco, Eckert_02, Tiesinga,
1dim, Bloch}.  During this overlap a phase that depends on the
two-qubits state, $|m,n\rangle \equiv |m\rangle\otimes|n\rangle$,
is accumulated because of the atom-atom (molecular) interaction.
Two-qubit phase gates based on this scheme were suggested
\cite{Brennen_99, Jaksch_99, Calarco}, analyzed \cite{Calarco,
Eckert_02, Tiesinga, 1dim}, and demonstrated \cite{Bloch}.

To achieve efficient computation and to be faster than decoherence
processes, it is desirable to generate gates which operate as fast
as possible \cite{Robin,phystoday}.  Cirac and Zoller
\cite{phystoday} discuss the slow two-qubit collisional gate as
one among two serious obstacles (the other is decoherence) to
quantum computing with atoms in optical potentials.  In designing
a fast collisional gate one is faced with the problem of leakage
outside the computational subspace due to rapid switching of the
control parameters.  A theoretical question with immediate
practical impact on the feasibility of quantum computation
emerges: how fast can such a gate be?  Previous theoretical work
used adiabatic evolution to ensure the fidelity of the gate.  In a
recent experimental implementation \cite{Bloch}, the time scale
for motion, 40 $\mu$s, was chosen ``to avoid any vibrational
excitations''.

Designing a fast two-qubit collisional gate is the purpose of this
paper.  We propose a fast-approach phase gate and use
optimal-control to implement it.  We first define the two-qubit
$\phi$ phase gate and explain its importance.  Next we review the
adiabatic realization of such gates in time-dependent optical
potentials.  Then we suggest the fast-approach scheme where
adiabaticity is not required.  Optimal control is applied and the
nature of the resulting dynamical path is analyzed.  Section
\ref{Two-Qubit Gate} describes two-qubit phase gates,
Sec.~\ref{The Model} sets out a simple model of a two-qubit phase
gate, and Sec.~\ref{Optimization of the Gate} presents the optimal
control scheme for optimizing the gate thereby implementing a fast
control phase gate.  In Sec.~\ref{Numerical Results} we present
numerical results of the optimization, Sec.~\ref{Further} applies
the fast-approach technique to improve the gate experimentally
demonstrated by \cite{Bloch}, and Sec.~\ref{Conclusion} concludes
the paper.  Appendix \ref{Entanglement power} shows that the phase
$\phi$ of a two-qubit phase gate uniquely the entanglement power
of the gate, and Appendix \ref{asi_pot} describes how to implement
a non-degenerate double well potential in an optical lattice.

\section{Two-Qubit Gate} \label{Two-Qubit Gate}

The two-qubit $\phi$ phase gate is designed to entangle two
interacting atoms (qubits) to a desired degree:
\begin{eqnarray}
&&{\rm P}(\phi)|m,n \rangle = \exp \left( {i\theta_{mn}} \right) |
m , n \rangle \,, \nonumber
\\
&&\theta_{00}+\theta_{11}-\theta_{01}-\theta_{10} \equiv  \phi~
({\rm mod} ~2\pi ) \,. \label{1}
\end{eqnarray}
This family of gates includes the control-phase gate:
\begin{equation}
{\rm CP}(\phi)|m,n \rangle =\exp\left({i m n \phi}\right)| m,n
   \rangle \,.
\end{equation}
Any P$(\phi)$ gate can be combined with unitary single-qubits Rabi
shifts, $\exp(i\alpha_m)$ $\exp(i\beta_n)$, to create the
CP$(\phi)$ gate, where $\theta_{mn}+\alpha_m+\beta_n = m n \phi\
({\rm mod} 2\pi )$, while
$\theta_{00}+\theta_{11}-\theta_{01}-\theta_{10}$ is invariant
under these shifts \cite{Tiesinga}.  The phase $\phi$ has an
intrinsic physical feature in that it parameterizes uniquely the
entanglement power of the gate (see Appendix A), and thus is
intimately connected with the coupling strength of the two atoms
during evolution.  The control phase gate with phase $\phi=\pi$
can further combine with single qubit operators to form the
control-not gate, ${\rm CN}|m,n \rangle=|m,n \oplus m \rangle $,
where $\oplus$ denotes addition modulo 2.  Each of these two-qubit
gates can be combined with a set of generators for single qubit
gates to form a universal set for quantum computation
\cite{universal-gates}.  In practice, a physical system may evolve
more naturally to a gate in P$(\pi)$ other than CN or CP$(\pi)$.
Therefore, in designing a gate, it is better to aim less
restrictively for any one of the equivalent P$(\phi)$ gates.

The basic idea of realizing phase gates with a two particle system
in an external potential is: The external potential initially
localizes the particles far enough apart so that they may be
considered independent.  The external potential then changes in
time so that wave function overlap gives rise to correlations due
to particle-particle interaction.  The external potential is
finally restored to its initial shape, so that the two particles
no longer interact, but are now in a new correlated state.

\section{The Model} \label{The Model}

As in Refs.~\cite{Calarco,Tiesinga,Eckert_02,Bloch,phystoday} we
focus on a two-qubit collisional gate where nearest-neighbor
interaction is manipulated by a time-dependent potential.  A
simple model for this process is a time-dependent Hamiltonian with
a double well potential, whose minima are separated by a
temporally varying distance $l(t)$:
\begin{eqnarray}
H &=& \frac{p_1^2}{2m}+\frac{p_2^2}{2m}+
V_1(x_1-l(t)/2) \nonumber \\
&+& V_2(x_2+l(t)/2) + 2\pi \hbar \omega_{\rm tr} a_s
\delta(x_1-x_2) \label{symham} ~.
\end{eqnarray}
In this model the ground and first-excited states of each trapped
atom form a single qubit computational basis.  We assume
distinguishable particles so no (anti)symmetrization is required.
The trapped particle interaction is modelled as an $s$-wave
scattering component of a Van der Walls interaction with
scattering length $a_s$ that reduces upon integration over the
transverse degrees of freedom to the above 1D interaction, and
$\omega_{\rm tr}$ is the frequency characterizing an harmonic
approximation for the transverse degrees of freedom
\cite{Eckert_02}.  To avoid problems due to degeneracy, the two
potential wells have different individual eigenvalues.  Otherwise,
when the coupling between degenerate qubits is switched on, there
will be fast oscillations between degenerate states regardless of
how slow the Hamiltonian changes in time.  Such a double well with
different frequencies can be obtained by two pairs of counter
propagating lasers with wave numbers $k,~ 2k$ and a bias electric
field $E$.  The lasers' relative phase and intensities determine
details of the double-well potential while the constant field can
be tuned so that the two minima have the same depth.  Varying $k$
by changing the angle between the incoming beams while increasing
$E$ and tuning the overall intensity can have the affect of
changing the distance between minima in time while keeping the
different trapping frequencies fixed.  In Appendix \ref{asi_pot}
we present additional details of an implementation of a
non-degenerate double well potential in an optical lattice.

When the two traps are at a distance $l$ apart, $E_{mn}(l)$ and
$|m, n; l \rangle$ are the instantaneous eigenvalues and
eigenvectors respectively: $H |m, n ; l \rangle = E_{m n}(l(t))|m,
n ; l \rangle$. Here $E_{mn}(l(t))=e_m^1+e_n^2+u_{mn}(l(t))$,
$e_m^1+e_n^2$ is the energy of the two atoms in their
non-interacting traps and $u_{mn}(l)$ is the interaction energy
which depends on the distance $l(t)$.  The asymptotic eigenstates
at $t=0$ and $\tau$, are direct-products of the individual trap
eigenstates, $|m \rangle = \Phi_m^1$ and $|n \rangle=\Phi_n^2$:
$|m, n \rangle \equiv |m, n ; l_0 \rangle = |m \rangle \otimes |n
\rangle$.  Such an initial eigenstate evolves into
\begin{equation}
{\cal U}(l;\tau,0)|m, n  \rangle=\sum_{m' n'} c^{m n}_{m' n'}(l)
|m', n' \rangle \,, \label{notlike1}
\end{equation}
at time $\tau$, where ${\cal U} (l;t_1,t_0)$ is the unitary
evolution from time $t_0$ to $t_1$ generated by Hamiltonian
(\ref{symham}) with a general time-dependent distance $l(t)$ and
the sum is over a complete Hilbert space.  A two-qubit gate is a
closed path $l(t)$, such that $l(0)=l(\tau)=l_0$ and the subspace
$W$ spanned by $\{ |00 \rangle,|01\rangle ,|10 \rangle, 11\rangle
\}$ is restored at time $t=\tau$, with high fidelity.  A
non-operating gate is one for which $l(t)=l_0$.  The distance
$l_0$ must therefore be such that the interaction energy for
states in $W$ can be neglected, i.e., $u_{mn}(l_0)\approx0$, for
$0\le n,m \le 1$.

\section{Optimization of the Gate} \label{Optimization of the Gate}

We wish to find a path $l(t)$, such that the restriction of ${\cal
U}(l;\tau,0)$ of Eq.~(\ref{notlike1}) on the computational
subspace $W$, ${\cal U}(l;\tau,0)|_W$, is equivalent to the
required gate of Eq.~(\ref{1}).  We do so by finding an objective
functional $J[l(t)]$, whose minimum is obtained when ${\cal
U}(l,\tau,0)|_W$ is equivalent to the required phase gate.  Given
a functional, $J[l]$, finding an optimum $l(t)$ reduces to well
established optimal-control functional analysis \cite{tannor}.  An
iterative procedure is applied; after solving for a given $l(t)$,
the functional $J[l]$ is evaluated and a gradient search method is
used to update $l(t)$ as a new, better, trial function.

In the case of adiabatic evolution, optimal control can be used to
enhance the fidelity of the gate, but is not essential.  In
adiabatic evolution, as long as no energy crossings are involved,
an eigenstate evolves to the same eigenstate, and
Eq.~(\ref{notlike1}) reduces to ${\cal U}(l;\tau,0) |m, n
\rangle=e^{{-\frac{i}{\hbar} \int_0^\tau E_{mn}(l(t)) dt}} | m, n
\rangle$.  Eq.~(\ref{1}) is trivially satisfied and ${\cal
U}(l;\tau,0)|_W$ is equivalent to a P$(\phi)$ phase gate with
$\phi=\int_0^\tau \Omega(l(t)) dt$, where $\Omega(l)$ is the
acquired controlled phase per unit time at distance $l$ given by:
\begin{equation}
\Omega(l)\equiv-\frac{1}{\hbar}\left(u_{00}(l)+u_{11}(l)-u_{01}(l)
- u_{10}(l) \right) \,.  \label{cpt}
\end{equation}
Note that $\Omega(l)$ is not monotonic and can have a maximal
value $\Omega_{\rm M}$ at $l_{\rm M}$ (e.g., see
Fig.~\ref{omeg_l}).

In designing a faster gate optimal control becomes essential.
Restoring the computational subspace after rapid switching of the
control parameters is not a trivial task.  After much iterative
work, using carefully chosen new objectives and numerical methods
as detailed below, we suggest the following fast-approach scheme:
{\em (a) Change $l(t)$ as fast as possible from $l_0$ to $l_{\rm
M}$, under the constraint that $\langle mn;l_{\rm M}| U(l;t(l_{\rm
M}),0) |m'n';l_0 \rangle \propto \delta_{mn}^{m'n'}$, i.e.
temporal eigenstates are restored.  (b) Let the atoms interact for
a time $\approx \phi / \Omega_{\rm M}$.  (c) Change $l(t)$ as fast
as possible from $l_{\rm M}$ back to $l_0$, such that the overall
evolution is diagonal in the computational basis.} This is similar
to previous suggestions for the phase gate with two differences:
the approach is not to the smallest possible distance but to the
optimized distance for acquiring an entanglement-effective phase,
and the approach and separation are not required to take place
adiabatically. The fundamental bound for the time required to
operate the gate is reduced to $\phi / \Omega_{\rm M}$ plus the
time required to evolve $l(t)$ from $l_0$ to $l_M$ and back.  As
shown below, the time required for the approach and descent is
reduced in this way by an order of magnitude, while high fidelity
is maintained.  Before presenting the optimal-control results we
comment on the choice of objective and the parameterization of
$l(t)$.

The choice of an objective out of the family of all equivalent
functionals is crucial.  One can use the objective $J_{\rm G}[l] =
-|\mbox{Tr}_W({\cal U}(l,\tau,0)G^\dagger)|$ whose minimal value
is obtained when ${\cal U}(l;\tau,0)|_W=G$ \cite{ronnie}. However,
to use this we would have to single out a specific gate $G\in {\rm
P} (\phi)$, whereas a physical system may evolve more naturally to
another.  Instead, we define a new objective $J_{\phi}[l]$ which
is minimal for {\em any} P$(\phi)$ gate:
\begin{equation}
J_{\phi}[l] \equiv |c_{00}\overline{c}_{01}\overline{c}_{10}c_{11}
- \exp({i\phi})| \,, \label{obj1}
\end{equation}
where $c_{mn} \equiv c^{mn}_{mn}$ and $c^{mn}_{m'n'}$ is defined
in Eq.~(\ref{notlike1}).  Similar objectives are given by
$J_{\phi}/F$ and $J_{\phi}/D$ where
\begin{equation}
F \equiv (1/4) \sum_{m,n,k,l=0}^1 |c_{mn}^{kl}|^2  \,,
\end{equation}
and
\begin{equation}
D \equiv (1/4) \sum_{m,n=0}^1 |c_{mn}|^2 \,,
\end{equation}
are fidelity measures.  $F$ quantifies the unitarity of the
evolution reduced to the subspace $W$.  This measure estimates how
close the evolution is to any two-qubit gate, and $1-F$ quantifies
leakage outside the computational basis.  The closer the second
fidelity measure $D$ is to unity, the closer the eigenstates
within $W$ evolve to themselves at time $\tau$.  A stationary path
for $J_{\phi}$ is necessarily a stationary maximal path for
$F,~D$, but not vice versa.

To solve the dynamics with Hamiltonian (\ref{symham}), we expand a
general solution $\zeta(t)$ in the basis of eigenstates
$\Psi_{mn}$ of the non-perturbed Hamiltonian $H_0$ ($a_s=0$):
\begin{eqnarray}
\zeta(t) = \sum_{mn} f_{mn} (t) \Psi_{mn}\exp
\left(-\frac{i}{\hbar}\left(e_m^1+e_n^2\right)t\right) \label{f}
\,, &&
\\
\Psi_{mn}(x_1,x_2,l) \equiv \Phi_m^1 \left( x_1-\frac{l}{2}
\right) \Phi_n^2 \left( x_2+\frac{l}{2} \right) \,. && \label{bas}
\end{eqnarray}
The Schr\"odinger equation reduces to
\begin{equation}
i\hbar\dot{\vec f \,}= \tilde U \vec f+\dot l \tilde M \vec f \,,
\label{schreq}
\end{equation}
where for an operator $X$, $\tilde X$ is short for $\tilde X
\equiv \exp \left({-i H_0 t/\hbar}\right) X \exp \left({i H_0
t}/{\hbar}\right)$, and
\begin{eqnarray}
&&U_{mn}^{kl} \equiv \gamma \int dx \, \Phi_m^1 (x_-) \Phi_n^2
(x_+) \Phi_k^1 (x_-) \Phi_l^2 (x_+)
\,, \\
&&M_{mn}^{kl}\equiv \langle \Psi_{mn}|p_2-p_1|\Psi_{kl} \rangle
\,, \label{defofm}
\end{eqnarray}
with $p_i\equiv-i\hbar\partial_{x_i}$, $\gamma=2\pi \hbar
\omega_{\rm tr} a_s$ and $x_\pm \equiv x \pm \frac{l}{2}$.  We
solved the system of differential equations using a stiff solver,
propagating four vectors of the subspace $W$.  The dimension of
the matrices was increased until the solution converged.  The
basis we chose is natural for adiabatic evolution so larger
matrices were required for the non-adiabatic simulations.

The trial functions for $l(t)$ were parameterized by:
\begin{eqnarray}
l=\Biggl\{ \begin{matrix} Q(q_1;\theta_1)~~~&\mbox{for}&~~~0\le t
\le t_1 \cr l_p~~~        &\mbox{for}&~~~t_1<<t<<t_2 \cr
Q(q_2;\theta_2)~~~&\mbox{for}&~~~t_2\le t \le \tau \end{matrix}
\end{eqnarray}
where $\theta_1 = {t}/{t_1}$ and $\theta_2 = ({\tau-t}) /
({\tau-t_2})$.  This describes a closed path $\{l(t): ~0\le t \le
\tau \}$.  For $t_1<<t<<t_2$, $l=l_p$ is fixed; we refer to this
as the plateau.  The approach and departure from the plateau are
characterized by $Q(q;x)\equiv l_0+(l_0-l_p)x^2(2x-3)+x^2(x-1)^2
q(x)$, where $q_i(x)$ are arbitrary polynomials.  With this
choice, both $l$ and $\dot l$ are continuous.  The coefficients of
$q_1$, $q_2$ along with $t_1$, $t_2$, $l_p$ (and sometimes $\tau$)
are adjusted by the optimal control scheme.  This parametrization
is suitable for both the adiabatic gate and the fast-approach
gate.

\section{Numerical Results} \label{Numerical Results}

\begin{figure}[htb]
\begin{center}
\centering
\includegraphics[scale=0.4,angle=0]{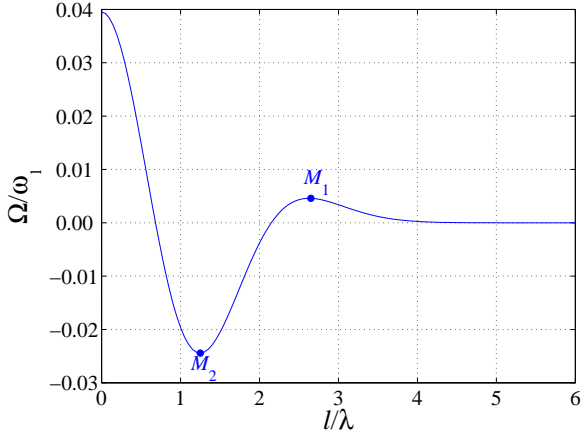}
\caption{The acquired controlled phase per unit time $\Omega$, as
a function of distance $l$.  The maximal value is at $M_2$ where
$l/\lambda=1.25$; a time of $128/\omega_1$ is required to obtain a
control-phase of value $\pi$.  } \label{omeg_l}
\end{center}
\end{figure}

In our numerical example the individual potential wells $i=1,2$
are harmonic with frequencies $\omega_i$.  The interaction
strength is then characterized by $\epsilon \equiv 2 \sqrt{\pi}
{a_s} \hbar \omega_{\rm tr} / {\lambda}$ where $\lambda^2 \equiv
\lambda_1^{2}+\lambda_2^{2}$ and $\lambda_i\equiv\sqrt{{\hbar}/{m
\omega_i}}$ is the harmonic oscillator length for well $i$.  (We
only consider $l>\lambda$ to avoid nonperturbative effects.)
Energy was taken in units of $\hbar \omega_1$, length in units of
$\lambda$, and time in units of $1/\omega_1$.  The interaction
strength was taken to be $\epsilon=0.05 \, \hbar\omega_1$,
corresponding to $ \omega_{\rm tr} \approx 10 \, \omega_1$, $a_s
\approx 0.005 \, \lambda$, and $\lambda_2= 1.2 \, \lambda_1$.  All
these numbers were chosen in the range of recent experiments in
optical lattices.  (E.g., for $^{87}$Rb, $a_s=0.005 \lambda$
corresponds to $\omega \approx 24$ Hz). Figure ~\ref{omeg_l} shows
$\Omega(l)$, the controlled phase acquired per unit time defined
in Eq.~(\ref{cpt}), where in leading order perturbation theory
$u_{mn}(l)=U_{mn}^{mn}$.  $\Omega(l)$ is non-monotonic with local
extrema at $M_1$ and $M_2$.  Most of the entanglement phase is
accumulated in our scheme during the plateau. Thus, despite the
phase lost on the way because the change in sign, we expect the
best plateau to be at $M_2$ where the control-phase acquired per
unit time is maximal.

\begin{figure}[htb]
\begin{center}
\centering
\includegraphics[scale=0.4,angle=0]{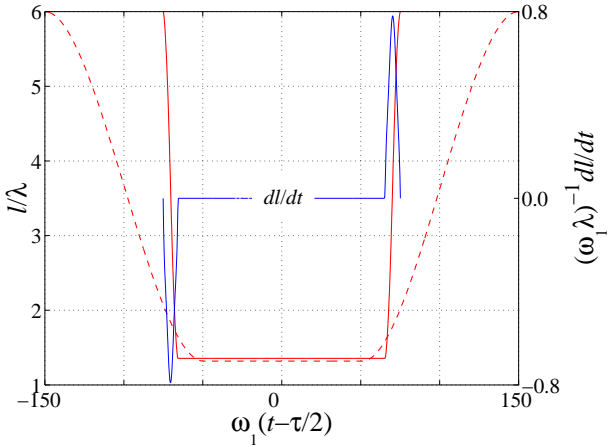}
\caption{The optimal path $l(t)$ obtained after optimization for
an adiabatic (dashed curve) and fast-approach (solid curve) gate
($\dot l$ also plotted for fast-approach). } \label{fig2}
\end{center}
\label{adi_l}
\end{figure}

\begin{figure}[htb]
\begin{center}
\centering
\includegraphics[scale=0.45,angle=0]{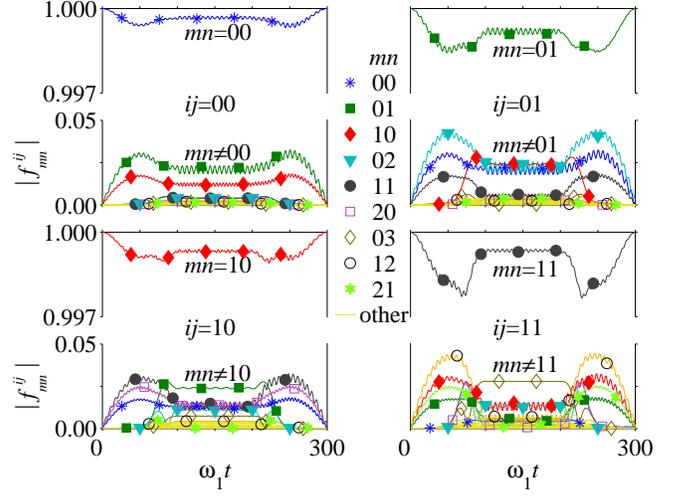}
\caption{ Absolute values of the coefficients $f_{mn}^{ij}(t)$
(Eq.(\ref{f})) for the optimal adiabatic gate.  The upper indexes
represent the four initial condition $f_{mn}^{ij}(0) = \delta_{im}
\delta_{jn}$, $0\le i,j \le 1$.  To a good approximation,
eigenstates are self-evolved at all times, as expected in
adiabatic evolution.}
\end{center}
\label{adi_coef}
\end{figure}

\begin{figure}[htb]
\begin{center}
\centering
\includegraphics[scale=0.45,angle=0]{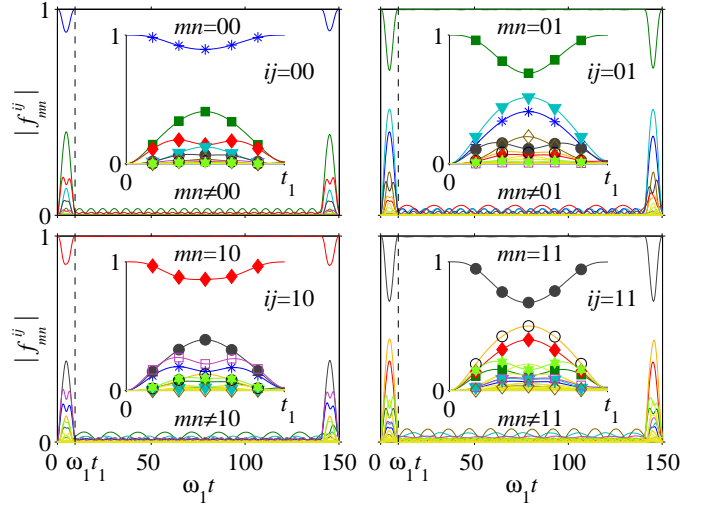}
\caption{ Absolute values of the coefficients $f_{mn}^{ij}(t)$, as
in Fig.~3, for the fast-approach gate.  Eigenstates in $W$ are
restored at the end of each fast step.}
\end{center}
\label{fa_gate}
\end{figure}

We first applied the optimization procedure to an adiabatic gate.
Good fidelity is expected in an adiabatic evolution.  However,
propagating adiabatically, one has an error of the order of $\dot
l$. Using the optimization scheme, the gate was improved and the
error reduced.  For the initial $l$ trial function we took
$l_p=l_{\rm M}=1.25 \lambda$, and $\tau=300 / \omega_1$,
$t_1=\tau-t_2=100 /\omega_1$, $q=0$.  This choice was constrained
by the requirement that the evolution be adiabatic, i.e., that
$|\dot l M| \ll 1$.  We expect some phase to be acquired outside
the plateau, so we took the plateau time to be smaller than
$\pi/|\Omega_{\rm_M}|\approx 128/\omega_1$.  The maximal value of
$|\dot l|$ is $\approx 0.07 {\lambda}/{\omega_1}$, consistent with
adiabaticity.  The optimized parameters are $l_p^{\rm opt}=
1.32\lambda =1.06 l_{\rm M}$, $t_1^{\rm opt}=100.47\omega_1$,
$t_2^{\rm opt}=200.94\omega_1$.  The optimized $l$ is shown in
Fig.~\ref{fig2}.  We ended the optimization with $J=0.00003$,
$F=0.99999 $, and $D=0.99999$.  Parameter variations of the order
of $10\%$ affect $D$ and $F$ to about one part in $10^5$, while
the acquired phase (here $\phi=3.1416$) is very sensitive to any
change in $l$, in particular ${\delta\phi}/{\phi} \approx (\delta
t_2 - \delta t_1)/(t_2-t_1)$, since $\phi\approx \Omega(l_p)
(t_2-t_1)$. The coefficients of the evolved computational basis
are shown in Fig.~3.

The fast-approach gate is considered next.  We expect the optimal
path for this scheme to be obtained when at the end of each fast
step (approach or departure) all eigenstates in $W$ are restored
to themselves while the required phase difference is obtained at
the plateau.  To select a trial function we therefore choose
$(t_2-t_1) = 128/ \omega_1$, $l_{\rm M}=1.25 \lambda$, $t_1 = 11
/\omega_1$, $\tau = 150 / \omega_1$.  This gives non-adiabatic
evolution, $\mbox{max} (| \dot l |) \approx 0.64 \,
\lambda/\omega_1$.  The optimized $l$ is the solid curve of
Fig.~\ref{fig2}.  The optimized parameters are $l_p^{opt}= 1.36 \,
\lambda$, $t_1^{opt}= 9.38 \, \omega_1$, $t_2^{opt}=140.49 \,
\omega_1$, $\tau^{opt} = 150.22 \, \omega_1$. Optimization
iterations were stopped when $J=0.006$, $D=0.997$, $F=0.999$, and
$\phi=3.14$.  The accuracy for $F$ and $D$ is maintained as long
as parameters for $l$ are $\approx 1\%$ from the optimized values.
With such variations, the accuracy for $\phi$ is $\approx 0.01$.
Varying the plateau time linearly from $0$ to
$t_2^{opt}-t_1^{opt}$ while keeping the approach and departure
shapes fixed, produces a linear curve for $\phi(t)$ with slope
$\Omega(l_p) = 0.0237 \omega_1$ with standard-deviation $\approx
10^{-4}$.  With this set of plateau times the fidelity measures
demonstrate oscillatory behavior about their optimized values with
deviations of about $10^{-3}$.  The coefficients of the evolved
computational basis are shown in Fig.~4.  Initial temporal
eigenstates were recovered at the plateau.  The approach and
departure times are reduced by a factor of $10$, leakage outside
the computational basis is restricted to less than 0.1\%, while
excellent agreement with an ideal P$(\pi)$ gate is maintained.

\section{A Further Application of the Fast-Approach Procedure}
\label{Further}

The fast-approach scheme suggested by our analysis is essentially
model independent.  For example, let us briefly consider the
implementation of such a fast-approach scheme for high-fidelity
transport in the experimental system of Refs.~\cite{Bloch} where
the computational basis was taken to be the hyperfine states of
$^{87}$Rb atoms trapped in an optical lattice in the Mott
insulating regime. The effective potentials experienced by the
atoms depended on their internal hyperfine states and the laser
polarization.  The gate was {\em not} implemented adiabatically.
For approach distances of half a lattice spacing and approach
times, $\tau$, longer than $\sim 2\pi/\omega$, where $\omega$ is
the harmonic frequency at the minimum of the optical lattice, the
degree of infidelity obtained was $\sim 5\%$.  In a preliminary
study, we applied optimal control to the approach step of this
gate.  The spatial wave packet of the atoms $\phi(x-R(t),t)$ can
be evolved upon moving the trap minima via a path $l(t)$ such that
$(\ddot R - \ddot l) + \omega^2 ( R - l) = - \ddot l$, and the
probability to escape the ground state at time $t$ is given by
$P(t) = 1 - |\langle\phi(x-l(t),t) | \phi(x-R(t),t)\rangle|^2 = 1
- e^{(-y^2(t)/4\lambda^2)}$, where $R(t)$ is the center of the
wave packet and $y(t) \equiv R(t)-l(t)$ with initial condition
$y(0)=0$ \cite{Japha}.  Taking $\sqrt{y^2(\tau)+\dot{y}^2(\tau)}$
as the objective, we were able to find a smooth path $l(t)$ that
reduces the approach time by 65\%, with $y^2(\tau) << 10^{-12}$ so
that losses were reduced from several percent to $P(\tau) \approx
2 y(\tau)^2 /4\lambda^2 \approx 10^{-12}$.  More work is planned
to minimize the run time, and implement the complete gate,
including the time interval of interaction, yet clearly our scheme
is able to significantly increase fidelity while somewhat reducing
the required time for the approach.

\section{Conclusion} \label{Conclusion}
In summary, using optimal-control, we improved the adiabatic
phase gate, designed a new fast-approach phase gate, and
demonstrated the new fast-approach scheme for a simple model and a
recent experiment. More sophisticated models with additional
degrees of freedom could further exploit optimal-control
techniques.

A few last remarks are in order.  The 1D model considered here is
obtained in reality from a physical 3D trap assuming that the
ground state of the transverse degrees of freedom is maintained
through the process.  A 3D model, where the moving potential
remains 1D, yet transverse excitations are allowed, could be
better for designing a fast gate.  In addition, for any scheme to
be practical, it is essential that the accuracy of the gate depend
weakly on small changes in $l$.  Here we saw that different
fidelity measures are quite robust, while the phase is very
sensitive to changes in $l$.  As a result the gates can be easily
adjusted to give a P$(\phi)$ gate with any $\phi$ by changing the
time duration of the plateau.  In practice, feedback learning
techniques may be used to assure stability and to adjust the
phase.  Finally, for a complete scheme of computation, one also
needs single qubit gates.  This is another point at which
degeneracy causes problems, as the computational basis of the two
oscillators needs to be un-entangled.  While these points deserve
further study, they were disregarded here, as our main focus was
that one could use optimal-control techniques and a fast-approach
scheme for the implementation of the control-phase gate in models
of cold atoms in an optical lattice, significantly reducing the
gates time while maintaining high fidelity.

{\it Acknowledgments} This research was supported by the Israel
Science Foundation (Grant Nos.~11-02 and 8006/03), U.S.-Israel
Binational Science Foundation (No.~2002147), and James Franck
Program (German-Israel) in Laser-Matter Interactions.  We are
grateful to David Tannor and Ehoud Pazy for useful discussions.

\appendix

\section{Entanglement power of the CP$(\phi)$ gate}
\label{Entanglement power}

An important property of the CP$(\phi)$ gate (and hence of any
P$(\phi)$ gate) is that the phase $\phi$ can be used to measure
the entanglement power of the gate.  To see this consider a
general two qubit state $|\Psi \rangle=\sum_{m,n=0}^1 a_{mn} |m,n
\rangle$. The action of CP$(\phi)$ on this state is,
\begin{equation}
A\equiv\begin{pmatrix} a_{00} & a_{01} \cr a_{10} & a_{11}
\end{pmatrix} ~\longrightarrow ~ \tilde A=\begin{pmatrix}
a_{00} & a_{01} \cr a_{10} & \sigma a_{11} \end{pmatrix}
\end{equation}
where $\sigma=\exp({i\phi})$.  Generally, if $\rho$ is the reduced
density matrix of a bipartite pure system, the usual entanglement
measures of $\rho$ are given by the purity (linear entanglement)
measure $M_1\equiv-{\rm Tr}{\rho^2}$, or the Von Neumann entropy
$M_2\equiv -\, {\rm Tr} \, \rho \log_2 \rho$.  There are important
mathematical features of interest when one chooses a measure, such
as convexity.  Here we are only interested in a measure of the
degree of entanglement.  The reduced density matrix of two qubits
can be written as
\begin{equation}
\rho= p {\cal P} + (1-p)(I - {\cal P}) \,,
\end{equation}
where $0\le p \le 1$, ${\cal P}$ is the one dimensional projection
onto the $p$ eigenstate of $\rho$, and $I$ is the identity in the
two level system sub-space.  For such a state
\begin{eqnarray}
&&M_1 = -[p^2 + (1-p)^2] \,,\\
&&M_2 = - p \log_2 p - (1-p) \log_2 (1-p)  \,.
\end{eqnarray}
The line ordering imposed by these measures is the same line
ordering obtained from the determinant of the reduced density
matrix: $\det(\rho)=p(p-1)$.  We therefore use the change in this
determinant as a measure for the entanglement induced by the gate:
\begin{eqnarray}
\Delta(\Psi;\phi) &\equiv& \mbox{det}(\tilde A \tilde A^\dagger)-
\mbox{det}(AA^\dagger) \\
&=& 2\Re\left(a_{00}\,a_{11}\,\bar
a_{10}\,\bar a_{01}\,(1-\sigma)\right) \\
&=& 4\Re\left(-i~r \exp\left({i (\gamma+\frac{\phi}{2})}\right)\,
\sin (\frac{\phi}{2})\right) \,,
\end{eqnarray}
where $r \exp({i\gamma})\equiv a_{00}\,a_{11}\,\bar a_{10} \, \bar
a_{01}$.  The maximal change is obtained when $r$ and
$\sin(\gamma+{\phi}/{2})$ are maximal, i.e., when
$|a_{mn}|={1}/{2}$ and $\gamma=({\pi-\phi})/{2}$. The entangling
power of the CP$(\phi)$ gate and hence of any P$(\phi)$ gate is
\begin{equation}
\mbox{max}_{\Psi}(\Delta(\Psi,\phi))=\frac{1}{4}\sin(\frac{\phi}{2})\,.
\end{equation}
Maximum entanglement power is obtained for $\phi=\pi$ (also for
the CN gate).

\section{Asymmetric double-well} \label{asi_pot}

Two counter-propagating laser beams at a fundamental frequency and at
its second harmonic, with wave numbers $k$ and $2k$, and intensities,
produce an effective standing wave optical potential of the form
\begin{equation}
V(k,x) = V_0 [\cos^2(k x+ \delta) + \alpha \cos^2(2 k x)],
\end{equation}
where $V_0$ is proportional to the intensity of the fundamental,
$\alpha$ is determined by beam intensities and detuning ratios, and
$\delta$ is determined by the relative phase between the fundamental
and second harmonic.  The wave number $k$ can be tuned by manipulating
the angle between the right and left laser beams.  The effective
potential thus formed is an array of double wells, with $\alpha$
determining the relative heights of the well minima.  For $\delta>0$,
the wells are non-symmetric.  Let $x_1(k)$, $x_2(k)$ denote the
positions of the minima of a specific double well, where $x_1(k)<
x_2(k)$.  The greater $\delta$, the larger $V(x_2(k))$ relative to
$V(x_1(k))$, as well as $V''(x_2(k))$ relative to $V''(x_1(k))$.  One
can add a linear potential of the form
\begin{equation}
\tilde V(x,k) = -A(k) x \label{v_t}
\end{equation}
to $V(k,x)$ so the resulting potential, $V_{\rm{eff}}$, will have equal
well minima, $V_{\rm{eff}}(x_2(k)) = V_{\rm{eff}}(x_1(k))$.  This
requires
\begin{equation}
 A(k) := V_0 \frac{V(x_2(k))-V(x_1(k))}{x_2(k)-x_1(k)} \,. \label{E}
\end{equation}
As a result of the additional linear potential, there is an increase
in the ratio of the second derivatives at the well minima of
$V_{\rm{eff}}$.  Note that the original potential $V(k,x)$ is a
function of $k x$, $V(k,x)=f(k x)$, thus
\begin{eqnarray}
V(k',x) &=& f(k \frac{k'}{k} x)=V(k,\frac{k'}{k} x) \,, \\
k' x_i(k') &=& k x_i(k) \,,
\end{eqnarray}
and hence
\begin{eqnarray}
\frac{A(k')}{k'} &=& \frac{V(k',x_2(k'))-V(k',
x_1(k'))}{k'\left( x_2(k')-x_1(k') \right)} \,, \\
 &=& \frac{ V(k,x_2(k))-V(k, x_1(k))}{k\left( x_2(k)-x_1(k)
\right)} \cr &=& \frac{A(k)}{k} \,.
\end{eqnarray}
$A(k)/k$ is independent of $k$ and (\ref{v_t}) can be written as
\begin{eqnarray}
\tilde V (k',x) =- V_0 C k' x
\end{eqnarray}
where $C$ is a constant that may be determined by (\ref{E}) for any
$k$.  Thus,
\begin{eqnarray}
V_{\rm{eff}}(k,x) = V_0 [ \cos^2(k x+ \delta) + \alpha \cos^2(2 k
x) - C k x] \,,
\end{eqnarray}
is a function of $k x$.  For such a potential, variations of $k$ scale
$x$.  The distance $l(k)$ between the two minima in a double well is
thus a parameter scaled by $k$, $l(k')= {k' \over k} l(k)$, while
keeping the values of potential values at minimal points unaltered as
well as the ratios of second derivatives at the minima.  To maintain
the values of second derivatives at the minimum constant as one varies
$k$, one can modify the laser intensities to depend on $k$ such that
\begin{equation}
    V_0 \rightarrow V_0(k) := \frac{V}{k^2} \,.
\end{equation}

\end{document}